\input mtexsis
\input epsf.tex
\paper
\singlespaced
\widenspacing
\twelvepoint
\Eurostyletrue
\sectionminspace=0.1\vsize
\def\discr{{\bf Z }_2} 
\def\nc{{N}_c}
\def\yo1{{{f_\pi}^2}}

\def\smallo{{\textstyle{1\over\sqrt{2}}}}

\def\oneh{ {1\over 2} }
\def\threeh{ {3\over 2} }
\def\oneht{\textstyle{1\over 2} }
\def\onehtsq{\textstyle{1\over{\sqrt{2}}} }

\def\sss{\scriptscriptstyle}

\def\discrete{ Z_2}
\referencelist
\reference{pdg} Particle Data Group, \journal Phys. Rev.D;54, (1996)
\endreference
\reference{ugo} U.~Aglietti, \journal Phys. Lett. B;281,341 (1992)
\endreference
\reference{avw} C.~Lovelace, \journal Phys. Lett. B;28,264 (1968)
\endreference
\reference{*avwa}  M.~Ademollo, G.~Veneziano, and S.~Weinberg, 
                  \journal Phys. Rev. Lett.;22,83 (1969)
\endreference
\reference{*avwb}  For a review, see S.~Weinberg, 
                  \journal Comm. Nucl. Part. Phys.;3,28 (1969)
\endreference
\reference{lew} D.~Lewellen, \journal Nucl. Phys. B;392,137 (1993)
\endreference
\reference{*lewa} J.~Polchinski, UTTG-16-92, {\tt hep-th/9210045}
\endreference
\reference{alg}  S.~Weinberg, \journal Phys. Rev.;177,2604 (1969)
\endreference
\reference{mended}  S.~Weinberg, \journal Phys. Rev. Lett.;65,1177 (1990);
           {\it ibid}, 1181
\endreference 
\reference{beane} S.R.~Beane, {{\tt hep-ph/9706246}}, Ann.Phys., in press
\endreference
\reference{wilson}  K.G.~Wilson and D.G.~Robertson, {{\tt hep-th/9411007}}
\endreference
\reference{*wilsona}  D.~Mustaki, {{\tt hep-ph/9410313}}
\endreference
\reference{*wilsonb} See also, L.~Susskind and M.~Burkardt, {{\tt hep-ph/9410313}}
\endreference
\reference{silas} S.R.~Beane, 
in {\it Proceedings of the AUP Workshop on QCD:
Confinement, Collisions and Chaos}, (World Scientific). p269,
{\tt hep-ph/9702268}; See also {\tt hep-ph/9512228}
\endreference
\reference{isospin}  S.~Weinberg, \journal Trans. New York Acad. Sci.II;38,185 (1977)
\endreference 
\reference{*isospina} For recent developments, see H.~Leutwyler, {\tt hep-ph/9602255}
\endreference 
\reference{sfsr}  S.~Weinberg, \journal Phys. Rev. Lett.;18,507 (1967)
\endreference
\reference{bura}  L.~Burakovsky and T.~Goldman, {\tt hep-ph/9703271}
\endreference
\reference{dashen} R.~Dashen, \journal Phys. Rev.;183,1245 (1969)
\endreference 
\reference{isgur} N.~Isgur and M.B.~Wise,  
                 \journal Phys. Rev. Lett.;66,1130 (1991)
\endreference
\reference{wise2} For a review, see M.B.~Wise, {\tt hep-ph/9311212}
\endreference
\reference{cley}  See, for instance, J.~Cleymans,
              \journal Nuov. Cim. A;4,433 (1971)
\endreference
\reference{nn}  H.B.~Nielsen and M.~Ninomiya
              \journal Nucl. Phys. B;185,20 (1981)
\endreference
\reference{ks}  L.H.~Karsten and J.~Smit, 
              \journal Nucl. Phys. B;183,103 (1981)
\endreference
\reference{sb2}  S.R.~Beane, {{\tt hep-ph/9706250}}
\endreference
\reference{sb3}  S.R.~Beane, in preparation
\endreference
\reference{wolo} H.R.~Fiebig and R.M.~Woloshyn, 
TRI-PP-96-1, {\tt hep-lat/9603001};\hfill\break
\journal Nucl. Phys. Proc. Suppl.;53,886 (1997) {\tt hep-lat/9607058}
\endreference

\endreferencelist
\titlepage
\singlespaced
\widenspacing

\obeylines
\hfill{DOE/ER/40762-123}
\hfill{U.ofMd.PP\#97-123}
\hfill{DUKE-TH-96-115}\unobeylines
\vskip0.5in
\title
Stringy Mass Squared Splittings Revisited
\endtitle
\author
Silas R.~Beane

Department of Physics, University of Maryland\Footnote\dag{Present Address.\hfill}
College Park, MD 20742-4111

{\it and}

Department of Physics, Duke University
Durham, NC 27708-0305

\vskip0.1in
\center{{\it sbeane@pion.umd.edu}}\endcenter
\endauthor
\abstract
\doublespaced
An empirically successful mass formula derived long ago from hadronic
string models is not explained by symmetries of the QCD lagrangian.
Consider that ${M_{\rho}^{2}}-{M_{\pi}^{2}} = {M_{\sss
{K*}}^{2}}-{M_{\sss K}^{2}}$ to a few percent accuracy. We derive this
and other equal spacing relations using chiral symmetry and a novel
$\discrete$ symmetry.  We offer an interpretation of this $\discrete$
symmetry as a manifestation of chiral anomalies in a gauge theory with
an intrinsic cutoff.
\endabstract 
\vskip0.5in
\center{PACS: 11.30.Rd; 12.38.Aw; 11.55.Jy; 11.30.Er} 
\endcenter
\endtitlepage
\vfill\eject                                     
\singlespaced
\widenspacing
\superrefsfalse

\vskip0.1in
\noindent {\twelvepoint{\bf 1.\quad Introduction}}
\vskip0.1in

Consider the meson mass-squared splittings listed in Table 1\ref{pdg}.
Why are these splittings equal to within a few percent?  The near
equality of the $D$-${D^*}$ and $B$-${B^*}$ mass-squared splittings is
well understood as a consequence of heavy quark symmetry. However,
equality of all other meson pairs is puzzling since there is no
obvious QCD symmetry which relates them.  The near equality of the
pairs could be accidental. There are rigorous mass formulas for heavy
hadrons that also happen to work well for light hadrons, providing a
similar mystery\ref{ugo}. However, the equal spacing relations of
Table 1 are special in that they are predicted by hadronic string
models made consistent with chiral symmetry constraints\ref{avw}. The
fact that the mass-squared splittings can be understood on the basis
of regularity, albeit regularity whose link with QCD is not clear,
suggests that they are not accidental. In the modern view, a
demonstration of these relations will be convincing only if there is a
symmetry argument based in QCD. The only post-sixties attempt to
understand these relations that we are aware of is in modern string
theory\ref{lew}.

In this paper we show that these relations can be understood using
$SU(N)_L\times SU(N)_R$ with $N=2,3$ together with a $\discrete$
symmetry that permutes chiral representations in a specific way. The
$\discrete$ symmetry completely determines the reducible chiral
representations filled out by mesons. The basic results for the chiral
representations have been found by Weinberg using algebraic sum
rules\ref{alg}\ref{mended}. Following \Ref{beane} we obtain Weinberg's
results directly using the $\discrete$ symmetry. What is new here is
the inclusion of explicit chiral symmetry breaking effects in the
mass-squared matrix, the extension of the basic multiplets to three
flavors of quarks and a discussion of the chiral representations of
the heavy mesons. Our derivation of the equal spacing relations is
related to the old string derivation; the constraints implied by the
full chiral algebra and the $\discrete$ symmetry are equivalent to
standard assumptions of Regge asymptotic behavior in pion-hadron
scattering\ref{alg}, which are automatically incorporated in hadronic
string models. What then does the $\discrete$ symmetry have to do with
QCD? We offer an interpretation of $\discrete$ in the context of
lattice QCD. We argue that $\discrete$ is a manifestation of chiral
anomalies in a gauge theory with an intrinsic cutoff.

\table{avw}
\caption{Lowest lying mesons of a given character. Masses are
central values from the particle data group\ref{pdg}, and we have
defined $\alpha ' \equiv{0.88}\;{{GeV}^{-2}}$ so as to normalize the
$\rho -\pi$ splitting to $0.50$. The ${\eta_8}$ and ${\phi_8}$ masses
are defined by the Gell-Mann-Okubo formulas (see below).}

\ruledtable
$A^*$-$A$          |  ${\alpha '}({M_{A^*}^2}-{M_{A}^2})$  \cr
$\rho -\pi$        |  $0.50$\crnorule
${K^*}-K$          |  $0.49$\crnorule
${\phi_8}-{\eta_8}$    |  $0.48$\cr
${D^*}-D$          |  $0.48$\crnorule
${B^*}-B$          |  $\;0.43$
\endruledtable
\endtable

This paper is organized as follows.  In Section 2 we discuss our basic
assumptions and use a $\discrete$ symmetry to obtain the chiral
representation of the Goldstone bosons.  We show that this
representation implies equal spacing relations for the Goldstone
bosons and their chiral partners. This is the most important Section
of the paper. Although Section 2.2 presents material that is well
known\ref{alg}\ref{mended}\ref{beane}, it is essential for what
follows. In Section 3 we consider the two-flavor chiral representation
involving $\eta$ and compare with Section 2 where $\eta$ is a
Goldstone boson of three-flavor QCD. Consistency of the two- and
three-flavor multiplets unambiguously determines members of the
$0^{++}$ scalar octet and requires octet-singlet mixing.  In Section 4
we use $\discrete$ to obtain the chiral representations of the $I=1/2$
mesons. We find the kaon representations and compare with the results
of Section 2 where the kaons are in the Goldstone boson representation
of three-flavor QCD.  This enables us to express the kaon mass-squared
splittings in terms of an $I=1/2$ matrix element, which we conjecture
to be universal.  We then construct the chiral representations of the
heavy mesons consistent with heavy quark symmetry, and show that the
heavy meson mass-squared splittings are related to the kaon splittings
by the universal $I=1/2$ matrix element. We discuss the baryons
briefly in Section 5 and provide an independent test of the
universality conjecture. In Section 6 we discuss the relation between
our derivation of the equal spacing relations and the old hadronic
string derivation. Finally, we give an interpretation of the
$\discrete$ symmetry in the context of lattice QCD. We summarize and
conclude in Section 7.

\vskip0.1in
\noindent {\twelvepoint{\bf 2.\quad The Goldstone Quartet}}
\vskip0.1in

\vskip0.1in
\noindent {\twelvepoint{\it 2.1\quad Mended Chiral Symmetry}}
\vskip0.1in

In a theory where a symmetry $G$ is spontaneously broken, $G$ is not a
symmetry of the vacuum, but $G$ is still a symmetry of the theory. The
noninvariance of the vacuum makes it difficult to recover consequences
of the symmetry in the broken phase. However, those consequences are
there and they are important.  Fortunately, in special Lorentz frames
the QCD vacuum can be invariant with respect to chiral symmetry even
when chiral symmetry is spontaneously broken\ref{alg}\ref{wilson}. The
infinite momentum frame provides an intuitive picture of how this can
occur. If a system is boosted to infinite momentum there is a sense in
which the vacuum decouples and is thereby rendered
irrelevant\ref{wilson}. In the infinite momentum frame helicity is
conserved and so, for each helicity, hadrons can be classified in
representations of the full chiral algebra in the broken
phase\ref{wilson}\ref{alg}. Since degeneracies in the broken phase are
uncommon, the chiral representations are generally reducible, and so
do allow non-trivial splitting of states. The symmetry constraints on
the hadronic mass-squared matrix will be the focus of this paper.  The
chiral representations also constrain amplitudes for pion emission and
absorption as discussed in many
places\ref{alg}\ref{mended}\ref{beane}\ref{silas}. Throughout this
paper it is assumed that we are working in Lorentz frames in which
helicity is conserved.

\vskip0.1in
\noindent {\twelvepoint{\it 2.2\quad $\discrete$ Multiplets}}
\vskip0.1in

Consider QCD with two massless flavors. There is an anomaly free
$SU(2)_L\times SU(2)_R$ symmetry which is assumed to be spontaneously
broken to $SU(2)_V$ (isospin).  We assume that the spectrum is
confined. Hadrons fall into isospin multiplets. For each helicity,
hadrons also fill out representations of $SU(2)\times
SU(2)$\ref{alg}. Since all mesons have zero helicity states, we will
consider only zero helicity in this paper.

Meson states carry isospin $0$ or $1$ and therefore transform as
combinations of $(2,2)$, $(3,1)$, $(1,3)$ and $(1,1)$ irreducible
representations of $SU({{{2}}})\times SU({{{2}}})$.  Charge
conjugation leaves $(2,2)$ and $(1,1)$ unchanged and interchanges
$(1,3)$ and $(3,1)$. Physical meson states have definite charge
conjugation and isospin and therefore are linear combinations of the
isovectors ${\ket{2,2}_a}$,
$\{{\ket{1,3}_a}-{\ket{3,1}_a}\}/{\textstyle{\sqrt{2}}}\equiv{\ket{V}_a}$
and
$\{{\ket{1,3}_a}+{\ket{3,1}_a}\}/{\textstyle{\sqrt{2}}}\equiv{\ket{A}_a}$,
and the isoscalars ${\ket{2,2}_4}$ and $\ket{1,1}$. Roman subscripts
are isospin indices. Only ${\ket{V}_a}$ changes sign under
charge conjugation.

In helicity conserving Lorentz frames the hadronic mass-squared
matrix, ${{\hat M}^2}$, is the natural object to
study\ref{alg}\ref{wilson}. We assume that the mass-squared matrix can
be written as ${{\hat M}^2}={{\hat M}_0^2}+{{\hat M}_{\sss \vev{{\bar
q}q}}^2}$.  This is the statement that all mass-squared splittings
between hadrons in a given chiral multiplet transform like the chiral
order parameter $\vev{{\bar q}q}$; i.e., like $(2,2)$ with respect to
$SU({{{2}}})\times SU({{{2}}})$.  ${{\hat M}_0^2}$ transforms like a
chiral singlet.  The only products of the allowed irreducible
representations that contain $(2,2)$ are $({2},{2})\otimes(3,1)$,
$({2},{2})\otimes(1,3)$, and $({2},{2})\otimes(1,1)$. It is clear that
a chiral representation with mass-squared splittings must be
reducible. All states in an irreducible representation must be
degenerate.

Consider a $\discrete$ symmetry that interchanges $(2,2)$ and
$(1,3),(3,1),(1,1)$ representations.  The permutations of meson states
consistent with isospin, charge conjugation and the allowed
mass-squared splittings are ${\ket{2,2}_a}\leftrightarrow{\ket{A}_a}$
and ${\ket{2,2}_4}\leftrightarrow\ket{1,1}$.  There are then two
reducible chiral representations consistent with the assumed
mass-squared splittings and the $\discrete$ symmetry$^1$\vfootnote1{In what
follows, $\mu^2$ and $\delta$ represent generic elements of the
mass-squared matrix which transform as $(1,1)$ and $(2,2)$,
respectively; that is, ${\mu^2} \in {{\hat M}_0^2}$ and $\delta \in
{{\hat M}_{\sss \vev{{\bar q}q}}^2}$.}:

\vskip0.1in
\noindent{\bf\underbar{a}}:\qquad 
$({2},{2})\otimes(3,1)\;{\it and}\; ({2},{2})\otimes(1,3)$
\vskip0.1in

\offparens
$$\eqalign{ 
&\ket{{\rm I}}_a=\smallo \{{\ket{2,2}_a}-{\ket{A}_a}\}
\hskip1.64in
{M_{\sss {I}}^2}={\mu^2}-\delta\cr
&\ket{{\rm II}}_a=\smallo \{{\ket{2,2}_a}+{\ket{A}_a}\}
\hskip1.575in
{M_{\sss {{II}}}^2}={\mu^2}+\delta\cr
&\ket{{\rm III}}={\ket{2,2}_4}\quad \ket{{\rm IV}}_a={\ket{V}_a}
\hskip1.404in
{M_{\sss III}^2}={M_{\sss {{IV}}}^2}={\mu^2}.\cr}
\EQN mes1$$ 
\autoparens
States $\ket{\rm I}$, $\ket{\rm II}$ and $\ket{\rm III}$ have charge
conjugation sign $\pm\epsilon$, $\ket{\rm IV}$ has sign $\mp\epsilon$.
States $\ket{\rm I}$ and $\ket{\rm II}$ form a $\discrete$ doublet in
the sense that the permutation
${\ket{2,2}_a}\leftrightarrow{\ket{A}_a}$ implies $(\ket{\rm
I},\;\ket{\rm II})\rightarrow (-\ket{\rm I},\;\ket{\rm II})$.
$\discrete$ is the discrete group whose elements are the square roots
of unity: $\{-1,1\}$.  In the right column we exhibit the mass
relations implied by the representation content. The lowest lying
member of this quartet must be an isovector.

\vskip0.1in
\noindent{\bf\underbar{b}}:\qquad $({2},{2})\otimes(1,1)$
\vskip0.1in

\offparens
$$\eqalign{ 
&\ket{{\rm I}}=\smallo \{{\ket{2,2}_4}-\ket{1,1}\}
\hskip2.025in {M_{\sss I}^2}={\mu^2}-\delta\cr
&\ket{{\rm II}}=\smallo \{{\ket{2,2}_4}+\ket{1,1}\}
\hskip1.95in {M_{\sss {II}}^2}={\mu^2}+\delta\cr
&\ket{{\rm III}}_a={\ket{2,2}_a}
\hskip2.73in{M_{\sss {{III}}}^2}={\mu^2}.\cr}
\EQN mes2$$ 
\autoparens
These states have the same charge conjugation sign. 
Again states $\ket{\rm I}$ and $\ket{\rm II}$ form a $\discrete$ doublet.
The lowest lying member of this triplet must be an isoscalar.
We will see examples of both $\bf a$ and $\bf b$ below.

Since the pion must be the lowest lying member of its chiral
representation, the pion must be in a representation of type $\bf a$.
In the case of zero-helicity {\it normality}, $\Pi\equiv P{(-1)^J}$,
is conserved, where $P$ is intrinsic parity and $J$ is spin. Since
$\pi$ has $\Pi$=$-1$, it follows that $\Pi_{\rm II}=-1$, $\Pi_{\rm
III} = \Pi_{\rm IV}= 1$.  Since $\Pi G$, where $G$ is $G$-parity,
commutes with the chiral algebra, the zero-helicity mesons fall into
distinct sectors labelled by $\Pi G$\ref{alg}. The pion has $\Pi
G$=$+1$ as do all states in its chiral representation. The quantum
number assignments are discussed in detail in
\Ref{alg}. Following \Ref{alg}, we will assume that the pion is joined in this
representation by a scalar $\epsilon$ ($\Pi$=$+1$), and the
zero-helicity components of $\rho$ ($\Pi$=$+1$) and ${a_{\sss 1}}$
($\Pi$=$-1$). We identify $\ket{\rm I}$$_a$=${ \ket{\pi}_a}$,
$\ket{\rm{II}}$$_a$ = $\ket{a_{\sss 1}}_a^{\sss (0)}$, $\ket{\rm{III}} =
\ket{\epsilon}$ and $\ket{\rm{IV}}$$_a$ = $\ket{\rho}_a^{\sss (0)}$. The
superscript denotes helicity. The chiral representation of the pion is
then

\offparens
$$\eqalign{ 
&\ket{\pi}_a=\smallo \{{\ket{2,2}_a}-{\ket{A}_a}\}\cr
&\ket{a_{\sss 1}}_a^{\sss (0)}= \smallo \{{\ket{2,2}_a}+{\ket{A}_a}\}\cr
&\ket{\epsilon}={\ket{2,2}_4}
\qquad\ket{\rho}_a^{\sss (0)}={\ket{V}_a}.\cr}
\EQN sfsr1$$ 
\autoparens
Here $\ket{\pi}$ and $\ket{a_{\sss 1}}^{\sss (0)}$ form a $\discrete$
doublet.  Sandwiching ${{\hat M}^2}$ between the states of \Eq{sfsr1}
gives

\offparens
$$\eqalign{
&{M_{\pi}^2}={{M}_0^2}-{{M}_{\sss \vev{{\bar q}q}}^2}\cr
&{M_{a_{\sss 1}}^2}={{M}_0^2}+{{M}_{\sss \vev{{\bar q}q}}^2}\cr
&{M_{\rho}^2}={M_{\epsilon}^2}={{M}_0^2}\cr}
\EQN ass1$$
\autoparens
where we have defined the matrix elements

\offparens
$$\EQNalign{
&\bra{2,2}\,{{\hat M}_0^2}\,\ket{2,2}=
\bra{A}\,{{\hat M}_0^2}\,\ket{A}=
\bra{V}\,{{\hat M}_0^2}\,\ket{V}\equiv{{M}_0^2} \EQN dfes;b\cr
&\bra{2,2}\,{{\hat M}_{\sss \vev{{\bar q}q}}^2}\,\ket{A}\equiv
{{M}_{\sss \vev{{\bar q}q}}^2}. \EQN dfes;b \cr}
$$\autoparens
It is useful to assign spurion transformation properties to the
mass-squared matrix elements. By definition, the mass-squared matrix
elements transform as ${{\hat M}_0^2} \rightarrow {{\hat M}_0^2}$ and
${{\hat M}_{\sss \vev{{\bar q}q}}^2} \rightarrow L{{\hat M}_{\sss
\vev{{\bar q}q}}^2}{R^\dagger}$ with respect to $SU({{{2}}})_L\times
SU({{{2}}})_R$. The $\discrete$ transformation
${{M}_0^2}\leftrightarrow{{M}_{\sss
\vev{{\bar q}q}}^2}$ implies $({M_{\pi}^2},\;{M_{a_{\sss 1}}^2})
\rightarrow(-{M_{\pi}^2},\;{M_{a_{\sss 1}}^2})$. 
Therefore, ${M_{\pi}^2}$ and ${M_{a_{\sss 1}}^2}$ transform as a
$\discrete$ doublet.  Of course in the chiral limit Goldstone's
theorem demands ${{M}_0^2}={{M}_{\sss \vev{{\bar q}q}}^2}$ and it
follows that ${M_{a_1}^2}=2{M_{\rho}^2}=2{M_{\epsilon}^2}$.

In summary, {\it a priori}, the chiral representations filled out by
light mesons in the broken phase are unknown. Irreducible
representations imply degeneracies that are uncommon in the hadron
spectrum.  On the other hand, reducible representations allow
mass-squared splittings and yet can be arbitrarily
complicated. However, a reasonable ansatz for the mass-squared matrix
together with a $\discrete$ symmetry reduces this {\it a priori}
infinite number of reducible chiral representations to two
representations with fixed mixing angles, which all light mesons with
nontrivial mass-squared splittings must fall into. The pion falls into
the unique reducible representation that allows massless isovector
states\ref{mended}\ref{beane}.

\vskip0.1in
\noindent {\twelvepoint{\it 2.3\quad Explicit Chiral Symmetry Breaking}}
\vskip0.1in

Assuming two degenerate flavors, to leading order in chiral
perturbation theory ${M_\pi^2}=2B{m_q}$, where ${m_q}={m_u}={m_d}$ and
$B\equiv -\vev{{\bar q}q}/2{f_\pi^2}$ with $\vev{{{\bar q}q}}=
\vev{{{\bar u}u}+{{\bar d}d}}$\ref{isospin}. We assign $B$ and $m_q$ the
$SU({{{2}}})_L\times SU({{{2}}})_R$ spurion transformation properties

\offparens
$$
\Gamma\rightarrow L\Gamma{R^\dagger}
\EQN spurtran $$\autoparens
where $\Gamma ={m_q}$ or $B$. These parameters
therefore transform as $(2,2)$. The product ${B{m_q}}$ transforms like
the quark mass operator in the QCD lagrangian and so should be an
invariant if spurion transformation properties have been properly
assigned.  In effect, since $(2,2)\otimes (2,2)$ contains the singlet
but not $(2,2)$, ${B{m_q}}\in{{\hat M}_0^2}$.  Explicit chiral
symmetry breaking effects are accounted for through the substitution
${M_0^2}\rightarrow{M_0^2}+2B{m_q}$ in \Eq{ass1}.  With
${{M}_0^2}={{M}_{\sss \vev{{\bar q}q}}^2}\equiv{\Delta}$ we then have

\offparens
$$\eqalign{
&{M_{\pi}^2}=2B{m_q}\cr
&{M_{a_{\sss 1}}^2}=2B{m_q}+2\Delta\cr
&{M_{\rho}^2}={M_{\epsilon}^2}=2B{m_q}+\Delta ,\cr}
\EQN ass1$$
\autoparens
from which follows the equal-spacing relation

\offparens
$$
{M_{\rho}^2}-{M_{\pi}^2}=
{M_{a_{\sss 1}}^2}-{M_{\rho}^2}=\Delta .
\EQN besame1$$
\autoparens
Note that these mass-squared splittings are independent of $O({m_q})$
explicit chiral symmetry breaking effects. This is so because the
quark mass contribution to the mass-squared matrix at leading order in
chiral perturbation theory is contained in ${{\hat M}_0^2}$ whereas
$\Delta\in{{\hat M}_{\sss
\vev{{\bar q}q}}^2}$, a consequence of the $\discrete$ symmetry
structure of the mass-squared matrix. This famous mass-squared
relation was originally obtained using spectral function sum
rules\ref{sfsr}. We will compare this relation to experiment in the
next section.

\vskip0.1in
\noindent {\twelvepoint{\it 2.4\quad Extension to $SU(3)$}}
\vskip0.1in

Consider QCD with three flavors. Mesons states transform as
combinations of $({\bar 3},3)$, $(3,{\bar 3})$, $(8,1)$, $(1,8)$ and
$(1,1)$ irreducible representations of $SU({{{3}}})_L\times
SU({{{3}}})_R$.  We now have the combinations
$\{{|{1,8})}-{|{8,1})}\}/{\textstyle{\sqrt{2}}}\equiv{|V)}$,
$\{{|{1,8})}+{|{8,1})}\}/{\textstyle{\sqrt{2}}}\equiv{|A)}$,
$\{{|{3,{\bar 3}})_{8}}-{|{{\bar 3},3})_{8}}\}
/{\textstyle{\sqrt{2}}}\equiv{|Y)_{8}}$ and $\{{|{3,{\bar
3}})_{8}}+{|{{\bar 3},3})_{8}}\}
/{\textstyle{\sqrt{2}}}\equiv{|X)_{8}}$.  The subscripts signify that
we are singling out the octet components of the $({\bar 3},3)\oplus
(3,{\bar 3})$ representations.  Only ${|V)}$ changes sign with
respect to charge conjugation.  The symmetry breaking mass-squared
matrix, ${{\hat M}_{\sss \vev{{\bar q}q}}^2}$, transforms like $({\bar
3},3)\oplus (3,{\bar 3})$ with respect to $SU({{{3}}})_L\times
SU({{{3}}})_R$.  Identical arguments as used above give the chiral
representations consistent with $\discrete$. The Goldstone boson
representation is\ref{mended}

\offparens
$$\eqalign{ 
&{|{\rm I})}=\smallo  \{ {|{X})_8}-{|{A})} \}\cr
&{|{\rm II})}=\smallo \{ {|{X})_8}+{|{A})} \}\cr
&{|{\rm III})}={|Y)_8}\qquad 
{|{\rm IV})}={|V)}.\cr}
\EQN mes1$$ 
\autoparens
We identify $|{\rm I}) = |{\cal P})$, $|{\rm II}) = |{\cal A})$,
$|{\rm III}) = |{\cal S})$ and $|{\rm IV}) = |{\cal V})$, where ${\cal
P}$ is the $0^{-+}$ Goldstone octet, ${\cal A}$ is the $1^{++}$
axialvector octet, ${\cal S}$ is the $0^{++}$ scalar octet and ${\cal
V}$ is the $1^{--}$ vector octet. Labelling by isospin as
$\{1,1/2,0\}$ we have ${\cal P}=\{\pi,K,{\eta_8}\}$, ${\cal V}= \{\rho
,{K^*}, {\phi_8}\}$, ${\cal S}=\{?,{K_0^*},? \}$ and ${\cal A}=
\{{a_{\sss 1}},{K_{1A}},{f_8}\}$\ref{pdg}. Several comments are in
order. The question marks refer to slots that are not unambiguously
filled by observed particles$^2$\vfootnote2{For a nice review of the
current situation, see \Ref{bura}.}.  The assignments are consistent
with the two-flavor chiral representation of \Eq{sfsr1} if we identify
$\epsilon_8$ as the $I=0$ member of the scalar octet. The physical
$\epsilon$ of the previous sections is then a mixture of $\epsilon_8$
with an $SU(3)$ singlet.  We postpone further discussion of the
scalars to the next section.  Generally, the $I=0$ members of the
octets mix with $SU(3)$ singlets when there is explicit symmetry
breaking. Specifically, $\eta_8$ mixes with the singlet $\eta_0$ to
give $\eta$ and $\eta '$, $\phi_8$ mixes with the singlet $\phi_0$ to
give $\omega$ and $\phi$ and $f_8$ mixes with the singlet $f_0$ to
give $f_{\sss 1}(1285)$ and $f_{\sss 1}(1510)$\ref{pdg}. We will
further discuss octet-singlet mixing below.  Following the particle
data group we treat ${K_{1A}}$ as an equal mixture of ${K_1}(1270)$
and ${K_1}(1400)$\ref{pdg}.

In extending to three flavors we assume that ${m_u}={m_d}\equiv
m\neq{m_s}$. It is straightforward to generalize our results for the
mass-squared matrix.  The mass-squared matrix elements are replaced
with the column vectors

\table{avwomesons}
\caption{The ${\phi_8}$ and ${f_8}$ masses 
are taken from the Gell-Mann-Okubo formulas.  ${K_{1A}}$ is assumed to
be an equal mixture of ${K_1}(1270)$ and ${K_1}(1400)$\ref{pdg}.  The
quoted uncertainties are due to the $a_{\sss 1}$, ${K_1}(1270)$ and
${K_1}(1400)$ masses.}

\ruledtable
$A^*$-$A$          |  ${\alpha '}({M_{A^*}^2}-{M_{A}^2})$  \cr
$a_{\sss 1}-\rho$  |  $0.81\pm 0.10$\crnorule
${K_{1A}}-{K^*}$   |  $0.86\pm 0.02$\crnorule
${f_8}-{\phi_8}$   |  $\,\, 0.90\pm 0.05$
\endruledtable
\endtable

\offparens
$$\eqalign{
&{{\hat M}_{\sss {\cal P}}^2}=
({M_\pi^2}\quad{M_K^2}\quad{M_\eta^2})^{\rm T} \cr
&{{\hat M}_{\sss {\cal V}}^2}=
({M_\rho^2}\quad{M_{K^*}^2}\;\;\,{M_{\phi_8}^2})^{\rm T} \cr
&{{\hat M}_{\sss {\cal A}}^2}=
({M_{a_{\sss 1}}^2}\;\;\,{M_{{K_{1A}}}^2}\;\,{M_{f_8}^2})^{\rm T}, \cr}
\EQN asskiss2$$
\autoparens
as is the quark mass:

\offparens
$$
{{\hat m}_q}=( m \quad (m+{m_s})/2 \quad (m+2{m_s})/3 )^{\rm T},
\EQN asskiss3$$\autoparens
whose elements are inferred from leading order chiral perturbation
theory\ref{isospin}.  We then have the generalization of \Eq{ass1}
to three flavors:

\offparens
$$\eqalign{
&{{\hat M}_{\sss {\cal P}}^2}=2B{{\hat m}_q}\cr
&{{\hat M}_{\sss {\cal V}}^2}=2B{{\hat m}_q}+\Delta{\bf 1}\cr
&{{\hat M}_{\sss {\cal A}}^2}=2B{{\hat m}_q}+2\Delta{\bf 1}\cr}
\EQN ass1sun$$
\autoparens
where $\bf 1$ is the
unit vector.  The Gell-Mann-Okubo formulas follow trivially from
\Eq{ass1sun}:

\offparens
$$\EQNalign{
&3{M_{\eta_8}^2}+{M_\pi^2}=4{M_K^2}\EQN askie;a\cr
&3{M_{\phi_8}^2}+{M_\rho^2}=4{M_{K^*}^2}\EQN askie;b \cr
&3{M_{f_8}^2}+{M_{a_{\sss 1}}^2}=4{M_{{K_{1A}}}^2}.\EQN askie;c \cr}
$$\autoparens 
We {\it define} $M_{\eta_8}$, $M_{\phi_8}$ and $M_{f_8}$ by the
Gell-Mann-Okubo formulas. From \Eq{ass1sun} follow also 
the equal spacing relations

\offparens
$$
{{\hat M}_{\sss {\cal V}}^2}-{{\hat M}_{\sss {\cal P}}^2}=
{{\hat M}_{\sss {\cal A}}^2}-{{\hat M}_{\sss {\cal V}}^2}=\Delta{\bf 1}
\EQN besamemucho$$
\autoparens
which imply

\offparens
$$\EQNalign{
&{M_\rho^2}-{M_\pi^2}={M_{K^*}^2}-{M_K^2}
={M_{\phi_8}^2}-{M_{\eta_8}^2}={\Delta}\EQN gpbia;a\cr
&{M_{a_{\sss 1}}^2}-{M_\rho^2}={M_{{K_{1A}}}^2}-{M_{K^*}^2}=
{M_{f_8}^2}-{M_{\phi_8}^2}={\Delta}.\EQN gpbia;b\cr}
$$
\autoparens
These equal spacing relations are the main result of this paper.
Clearly not all Gell-Mann-Okubo and equal-spacing relations are
independent. For instance \Eq{askie;a}, \Eq{askie;b} and \Eq{gpbia;a}
together comprise four relations, three of which are independent. 

The ${\cal V}-{\cal P}$ equal spacing rule works remarkably well (see
Table 1). The ${\cal A}-{\cal V}$ equal spacing rule is consistent
within the error bars (see Table 2). However, the equality of
\Eq{gpbia;a} and \Eq{gpbia;b} is not very good. 
The reason underlying this combination of 
remarkable accuracy and mediocrity is
mysterious. But we emphasize that the lack of agreement between
\Eq{gpbia;a} and \Eq{gpbia;b}
is nothing new; \Eq{gpbia} is a generalization of the famous relation,
\Eq{besame1}, familiar from spectral function sum rules\ref{sfsr}.
For instance, using \Eq{besame1} to predict the ${a_{\sss 1}}$ mass
from the $\rho$ and $\pi$ masses gives ${M_{a_{\sss 1}}}=1080MeV$
compared with the measured value of $1260\pm 40MeV$.

\vskip0.1in
\noindent {\twelvepoint{\it 2.5\quad Isospin Violation and Current Quark Masses}}
\vskip0.1in

\table{avwisovio}
\caption{Charged and neutral kaon mass-squared splittings from the
particle data group\ref{pdg}. We again use
$\alpha ' \equiv{0.88}\;{{GeV}^{-2}}$.}
\ruledtable
$A^*$-$A$          |  ${\alpha '}({M_{A^*}^2}-{M_{A}^2})$  \cr
${K^{*+}}-{K^+}$    |  $0.4851\pm 0.0004$\crnorule
${K^{*0}}-{K^0}$   |  $\; 0.4886\pm 0.0004$
\endruledtable
\endtable

We now consider ${m_u}\neq{m_d}$, to leading order in chiral
perturbation theory, and include an electromagnetic mass,
$\Delta{m_\gamma^2}$, consistent with Dashen's theorem\ref{dashen} in
the elements of ${{\hat M}_0^2}$ that carry electromagnetic charge. We
can thus test the universality of the electromagnetic corrections in
the ${\cal V}-{\cal P}$ equal spacing rule. We find

\offparens
$$\EQNalign{
&{M_{K^{*+}}^2}-{M_{K^+}^2}={M_{K^{*0}}^2}-{M_{K^0}^2}\EQN isovio;a\cr
&{M_{\rho^{+}}^2}-{M_{\pi^{+}}^2}={M_{\rho^{0}}^2}-{M_{\pi^{0}}^2}.
\EQN isovio;b\cr}
$$\autoparens 
\Eq{isovio;b} is consistent within error bars. \Eq{isovio;a} is compared
with experiment in Table 3.  The charged and neutral kaon splittings
do not agree within experimental errors.  Therefore this equal-spacing
rule distinguishes isospin violating contributions to the ${\cal P}$
and ${\cal V}$ octets. It is therefore of interest to calculate ratios
of quark masses using ${\cal V}$. We have

$$
{{m_u}\over{m_d}}=
{{{M_{K^{*+}}^2}-{M_{K^{*0}}^2}+2{M_{\pi^{0}}^2}-{M_{\pi^{+}}^2}}\over
{{M_{K^{*0}}^2}-{{M_{K^{*+}}^2}+{M_{\pi^{+}}^2}}}}=0.33\pm 0.05
\EQN mumd
$$
which is related by \Eq{isovio;a} to the usual relation involving
$\cal P$ alone which gives ${{m_u}/{m_d}}=0.55$\ref{isospin}.  The
errors are due to the $K^*$ masses\ref{pdg}. We also find

$$
{{m_s}\over{m_d}}=
{{{M_{K^{*0}}^2}-{M_{K^{*+}}^2}+2{M_{K^{+}}^2}-{M_{\pi^{+}}^2}}\over
{{M_{K^{*0}}^2}-{{M_{K^{*+}}^2}+{M_{\pi^{+}}^2}}}}=16.7\pm 0.6
\EQN msmd
$$
which is again related by \Eq{isovio;a} to the usual relation
involving $\cal P$ alone which gives ${{m_s}/{m_d}}=20.1$. The values
of the quark mass ratios implied by $\discrete$ and chiral symmetry
are not at odds with the usual predictions implied by chiral symmetry
alone if one takes into account theoretical error due to omitted
higher orders in the chiral expansion\ref{pdg}.
\vskip0.1in
\noindent {\twelvepoint{\bf 3.\quad The Eta Triplet and the Scalar Octet}}
\vskip0.1in

Hadrons participating in representations of $SU(3)\times SU(3)$ must
also participate in representations of $SU(2)\times SU(2)$. That the
representations be compatible is a nontrivial consistency check and
provides insight into the nature of octet-singlet mixing.  We saw
above that consistency of the pion representations required that
$\epsilon_8$ be the $I=0$ member of the scalar octet. Consistency of
the kaon $I=1/2$ representation with the Goldstone boson
representation will prove useful in relating matrix elements of light
mesons to those of heavy mesons in the heavy quark limit, as we will
see in the next section.  Here we discuss the lowest lying isoscalar,
$\eta$. Assume $\eta$ is in an irreducible representation of
$SU(2)\times SU(2)$. If $\eta$ is an $SU(2)\times SU(2)$ singlet then
it does not communicate with other states by pion emission and
absorption. This is in contradiction with the $SU(3)\times SU(3)$
assignment and is therefore ruled out. If $\eta$ is in a nontrivial
irreducible representation, then it must be degenerate with at least
an isovector other than $\pi$.  This is again in contradiction with
the $SU(3)\times SU(3)$ assignments.  Therefore, $\eta$ must be in a
reducible representation of type $\bf b$, found in Section 2.2.

\table{etatriplet}
\caption{Mass-squared splittings for the $\eta$ triplet from the
particle data group\ref{pdg}.} 
\ruledtable
$A^*$-$A$          |  ${\alpha '}({M_{A^*}^2}-{M_{A}^2})$  \cr
${f_{\sss 1}}-{a_{\sss 0}}$ | $0.59\quad $\crnorule
${a_{\sss 0}}-\eta$ |   $\;\, 0.59\quad $
\endruledtable
\endtable

Since $\eta$ has $\Pi$=$-1$, it follows that $\Pi_{\rm II}=-1$,
$\Pi_{\rm III} = 1$. All states have positive charge conjugation sign.
The $\eta$ representation is labelled by $\Pi G$=$-1$. Consultation of
the particle data tables\ref{pdg} makes clear that $\eta$ must be
joined in this representation by a scalar (isovector) ${a_{\sss
0}}(980)$ ($\Pi$=$+1$), and the zero-helicity component of ${f_{\sss
1}}(1285)$ ($\Pi$=$-1$).  The next candidate $I=1$ state with
appropriate quantum numbers to participate in the $\eta$ triplet is
the recently discovered ${a_{\sss 0}}(1450)$\ref{pdg}.  However, this
state, being heavier than ${f_{\sss 1}}(1285)$, cannot participate in
the $\eta$ representation.  We therefore identify $\ket{\rm I} =
\ket{\eta}$, $\ket{\rm{II}} = \ket{f_{\sss 1}}^{\sss (0)}$ and
$\ket{\rm{III}}$$_{a}$ =
$\ket{{a_{\sss 0}}}$$_{a}$. The chiral representation of $\eta$ is
then

\offparens
$$\eqalign{ 
&\ket{\eta}=\smallo \{{\ket{2,2}_4}-\ket{1,1}\}\cr
&\ket{f_{\sss 1}}^{\sss (0)}=\smallo \{{\ket{2,2}_4}+\ket{1,1}\}\cr
&{\ket{{a_{\sss 0}}}_{a}}={\ket{2,2}_a}.\cr}
\EQN meswer2$$ 
\autoparens
It follows that 

\offparens
$$\eqalign{
&{M_{\eta}^2}={{\bar M}_0^2}-{{\bar M}_{\sss \vev{{\bar q}q}}^2}\cr
&{M_{f_{\sss 1}}^2}=
{{\bar M}_0^2}+{{\bar M}_{\sss \vev{{\bar q}q}}^2}\cr
&{M_{a_{\sss 0}}^2}={{\bar M}_0^2}\cr}
\EQN triplet1$$
\autoparens
where we have defined the matrix elements

\offparens
$$\EQNalign{
&\bra{2,2}\,{{\hat M}_0^2}\,\ket{2,2}=
\bra{1,1}\,{{\hat M}_0^2}\,\ket{1,1}\equiv{{\bar M}_0^2}\EQN dog;a\cr
&\bra{2,2}\,{{\hat M}_{\sss \vev{{\bar q}q}}^2}\,\ket{1,1}\equiv
{{\bar M}_{\sss \vev{{\bar q}q}}^2}. \EQN dog;b \cr}
$$\autoparens
From \Eq{triplet1} follows the equal-spacing relation

\offparens
$$
{M_{f_{\sss 1}}^2}-{M_{a_{\sss 0}}^2}={M_{a_{\sss 0}}^2}-{M_{\eta}^2}
\EQN tripleteqsp
$$\autoparens which is compared with experiment in Table 4. This
relation works remarkably well.  Of course in two-flavor QCD there is
no reason why these splittings should be related to those involving
the pion quartet, \Eq{besame1}. However, in three-flavor QCD $\eta_8$
is in the pseudoscalar octet of Goldstone bosons and $f_8$ is in the
axialvector octet. Consistency requires that we treat ${a_{\sss 0}}$
as the $I=1$ member of the scalar octet. Noting that
${{\hat M}_{\sss {\cal S}}^2}={{\hat M}_{\sss {\cal V}}^2}$
we find

\offparens
$$\eqalign{
&{M_{\eta_8}^2}=2B(m +2{m_s})/3\cr
&{M_{f_8}^2}=2B(m +2{m_s})/3+2{\Delta}\cr
&{M_{a_{\sss 0}}^2}=2Bm+{\Delta},\cr}
\EQN triplet1$$
\autoparens
which when combined with \Eq{tripleteqsp} gives

\offparens
$$({M_{\eta_8}^2}-{M_{\eta}^2})+({M_{f_8}^2}-{M_{f_{\sss 1}}^2})=
{\textstyle {8\over 3}}({M_{K}^2}-{M_{\pi}^2}).
\EQN mixingsreq
$$\autoparens Notice that consistency of the two- and three-flavor
chiral representations {\it requires} nontrivial mixings in the
presence of explicit $SU(3)$ breaking effects. Therefore, without
octet-singlet mixing the two- and three-flavor chiral representations
cannot be made compatible.  We can use this equation together with the
Gell-Mann-Okubo formulas to predict ${M_{{K_{1A}}}}=1315MeV$ which is
consistent with the value $1340MeV$ which follows from assuming
$K_{1A}$ to be an equal mixture of ${K_1}(1270)$ and
${K_1}(1400)$\ref{pdg}.

Similar considerations apply to the $I=0$ member of the scalar
octet. Using ${M_{\epsilon_8}^2}=2B(m +2{m_s})/3+{\Delta}$ and
${M_{\epsilon}^2}=2Bm+{\Delta}$ we find

\offparens
$$({M_{\epsilon_8}^2}-{M_{\epsilon}^2})=
{\textstyle {4\over 3}}({M_{K}^2}-{M_{\pi}^2}).
\EQN mixingsreqep
$$\autoparens Presumably $\epsilon$, a mixture of
$\epsilon_8$ with an $SU(3)$ singlet, is identified with $f_{\sss
0}(400-1200)$\ref{pdg}.  Combining \Eq{mixingsreq} and
\Eq{mixingsreqep} gives the remarkable equation

\offparens
$$({M_{\eta_8}^2}-{M_{\eta}^2})+({M_{f_8}^2}-{M_{f_{\sss 1}}^2})=
{2}({M_{\epsilon_8}^2}-{M_{\epsilon}^2}).
\EQN mixingsreq
$$\autoparens 
We emphasize that this relation is a consequence of chiral
symmetry and $\discrete$.

In summary, we have found the two-flavor reducible chiral
representation filled out by $\eta$ and its chiral
partners. Consistency with the three-flavor representation is achieved
only if there is octet-singlet mixing. Similar considerations apply to
$\epsilon$.  We conclude on the basis of the consistency of the two-
and three-flavor chiral representations of $\pi$ and $\eta$ that the
scalar octet is ${\cal S}=\{{a_{\sss 0}}(980),{K_0^*},{\epsilon_8}\}$.
Consistency conditions for the kaons require discussion of $I=1/2$
chiral multiplets consistent with $\discrete$.

\vskip0.1in
\noindent {\twelvepoint{\bf 4.\quad Isospinor Multiplets}}
\vskip0.1in

\vskip0.1in
\noindent {\twelvepoint{\it 4.1\quad Isospinor $\discr$ Doublets}}
\vskip0.1in

The only representations of $SU(2)\times SU(2)$ that contain only a
single $I={\oneh}$ representation of the diagonal isospin subgroup are
$(0,{{\oneh}})$ and $({{\oneh}},0)$$^3$\vfootnote3{ Note that here we
label states by their isospin rather than the number of independent
components. For instance, in terms or our previous notation we have
$(0,{{\oneh}})=(1,2)$.}. So, in general, $I={{\oneh}}$ meson states of
definite helicity are linear combinations of any number of these
irreducible representations with undetermined
coefficients\ref{mended}. Mass splitting can only occur as a
consequence of mixing between these representations since ${\hat M}^2$
is a sum of $(0,0)$ (${{\hat M}_0^2}$) and $({{\oneh}},{{\oneh}})$
$({{\hat M}_{\sss \vev{{\bar q}q}}^2})$ contributions.  Consider two
$(0,{{\oneh}})$ representations, distinguished by a bar.  Then
$(\overline{0,{{\oneh}}})\otimes ({{\oneh}},0) =(2,2)$ and
$(\overline{0,{{\oneh}}})\otimes (0,{{\oneh}})=(1,1\oplus
3)$. Therefore at the level of the $I=1/2$ states $\discrete$ is the
permutation $(0,{{\oneh}})\leftrightarrow ({{\oneh}},0)$, which is
consistent with the permutation $(2,2)\leftrightarrow (1,1\oplus 3)$
which determined the chiral representations of the $I=0,1$ mesons.
$\discrete$ implies the chiral representation:

$$\eqalign{
&\ket{\rm I}=\onehtsq\{\ket{0\; {\oneht}}
    -\ket{{\oneht}\; 0}\}\hskip2.025in {M_{\sss I}^2}={\mu^2}-\delta\cr        
&\ket{\rm II}=\onehtsq\{\ket{0\; {\oneht}}
+\ket{{\oneht}\; 0}\}\hskip1.97in {M_{\sss II}^2}={\mu^2}+\delta\cr}
\EQN basicmultst $$\autoparens
where $\ket{\rm I}$ and $\ket{\rm II}$ are a $\discrete$
doublet\ref{mended}. In the right column we exhibit the mass relations
implied by the representation content. It also follows from
\Eq{basicmultst} that $\bra{\rm I}$${X_a}$$\ket{\rm II}$=${\tau_a}/2$
where $X_a$ is related to the amplitude for pion transitions and
$\tau_a$ are the Pauli matrices\ref{mended}. States $\ket{\rm I}$ and
$\ket{\rm II}$ experience no pion transitions to other states.

It is possible to build more complicated representations out of the
basic multiplet of \Eq{basicmultst} which are consistent with
$\discrete$. However, the mass-squared splittings and pion transitions
will always be equivalent to what is implied by the basic multiplet of
\Eq{basicmultst}.  We will see an example of this below.

\vskip0.1in
\noindent {\twelvepoint{\it 4.2\quad $K$-$K^*$ and ${K_0^*}-{K_{1A}}$ Consistency}}
\vskip0.1in

Since the kaons carry $I=1/2$ they must fall into $SU(2)\times SU(2)$
representations as well as the $SU(3)\times SU(3)$ representation
found in Section 2.  On the basis of our assumptions about the
mass-squared matrix ${K}$ and ${K^*}$ must either be paired with each
other in the sense of \Eq{basicmultst} or with other states.  If
${K^*}$ and ${K}$ are not paired with each other then they do not
communicate by single pion emission and absorption. But this cannot
be. The $SU(3)\times SU(3)$ representation of \Eq{mes1} implies that
$|\bra{K}{X_a}\ket{{K^*}}|\neq 0$. In particular, it implies
$|\bra{K}|X|\ket{{K^*}}|=|\bra{\pi}|X|\ket{{\rho}}|$ for reduced
matrix elements in the chiral limit. Since the latter does not vanish
in any limit, neither does the former.  Therefore ${K^*}$ and ${K}$
must be paired as

$$\eqalign{
&\ket{K^*}^{\sss (0)}=\onehtsq\{\ket{0\; {\oneht}\; s}
           +\ket{{\oneht}\; 0\; s}\}\cr 
&\ket{K}=\onehtsq\{\ket{0\; {\oneht}\; s}-\ket{{\oneht}\; 0\; s}\}\cr}
\EQN chstates3 $$\autoparens
where $s$ denotes strange quark content.  One can check that the
$SU(3)\times SU(3)$ and $SU(2)\times SU(2)$ predictions for the pion
transition matrix element, $\bra{K}{X_a}\ket{{K^*}}$, are
consistent\ref{silas}.  It is straightforward to find

$${M_{K^*}^2}-{M_{K}^2}= 
2\bra{{\oneht}\; 0\; s}{{\hat M}_{\sss \vev{{\bar q}q}}^2}
\ket{0\;{\oneht}\; s}. \EQN kmr$$ 
Consistency with the $SU(3)\times SU(3)$ 
result of \Eq{gpbia;a} then implies

$$\bra{{\oneht}\; 0\; s}{{\hat M}_{\sss \vev{{\bar q}q}}^2}
\ket{0\;{\oneht}\; s} =
\bra{{\oneht}\; 0}{{\hat M}_{\sss \vev{{\bar q}q}}^2}
\ket{0\;{\oneht}} =\oneht\Delta \EQN kmr2$$ 
where in the second step we have removed the $s$ label since {\it this
matrix element does not depend on strange quark properties}.  Of
course we have only shown this to be the case to leading order in
chiral perturbation theory. 

Similarly, $K_0^*$ and $K_{1A}$ must be paired as

$$\eqalign{
&\ket{K_{1A}}^{\sss (0)}=\onehtsq\{\ket{\overline{0\; {\oneht}\; s}}
           +\ket{\overline{{\oneht}\; 0\; s}}\}\cr 
&\ket{K_0^*}=\onehtsq\{\ket{\overline{0\; {\oneht}\; s}}
-\ket{\overline{{\oneht}\; 0\; s}}\}\cr}
\EQN chstatesk1a $$\autoparens
where the bar denotes that the states are distinct from
those of the $K-{K^*}$ pair. It follows that

$${M_{K_{1A}}^2}-{M_{K_0^*}^2}= 
2\bra{\overline{{\oneht}\; 0\; s}}{{\hat M}_{\sss \vev{{\bar q}q}}^2}
\ket{\overline{0\;{\oneht}\; s}}. \EQN cestlar$$ 
Again we find the consistency condition

$$\bra{\overline{{\oneht}\; 0\; s}}{{\hat M}_{\sss \vev{{\bar q}q}}^2}
\ket{\overline{0\;{\oneht}\; s}} =
\bra{\overline{{\oneht}\; 0}}{{\hat M}_{\sss \vev{{\bar q}q}}^2}
\ket{\overline{0\;{\oneht}}} =\oneht\Delta . \EQN kmr2asd$$ 

Given \Eq{kmr2} and \Eq{kmr2asd} we will assume that there is a {\it
universal} matrix element in the sense that

$$\bra{{\oneht}\; 0\; {\cal X}}{{\hat M}_{\sss \vev{{\bar q}q}}^2}
\ket{0\;{\oneht}\; {\cal X}} =\oneht\Delta \EQN kmr234$$ 
where ${\cal X}$ represents other quantum numbers carried by the
$I=1/2$ state. This additional assumption will prove essential in
relating the light- and heavy-meson mass-squared splittings. 

It might appear that the heavy meson pairs, $D$-${D^*}$ and
$B$-${B^*}$, should be paired like $K$-${K^*}$ in \Eq{chstates3}.  We
will see that such a simplistic assignment would violate constraints
due to heavy quark symmetry.

\vskip0.1in
\noindent {\twelvepoint{\it 4.3\quad The Heavy Mesons}}
\vskip0.1in

Heavy quark symmetry places constraints of its own on the heavy meson
mass matrix.  The ground state heavy mesons have light-quark spin
${s_{\ell}}={\oneh}$ and ${\pi _{\ell}}=(-)$ and are denoted $P$
($0^-$) and $P^*$ ($1^-$)\ref{isgur} where $P$ is $D$ or $B$. Their
masses can be expressed as

\offparens
$$
{M_P}={m_Q}+{\bar\Lambda}
+\{{\tilde K}+{\tilde G}\}/{m_Q}+O({1/{m_Q^2}})
\EQN qcdmasssq1$$\autoparens
\offparens
$$
{M_{P^*}}={m_Q}+{\bar\Lambda}
+\{{\tilde K}-{\textstyle {1\over 3}}{\tilde G}\}/{m_Q}
+O({1/{m_Q^2}}),
\EQN qcdmasssq2$$\autoparens
where ${\bar\Lambda}$ is a positive contribution ---independent of the
heavy quark mass--- and ${\tilde K}$ and ${\tilde G}$ are matrix
elements of heavy quark operators which are also independent of the
heavy quark mass\ref{wise2}. Squaring the masses gives

\offparens
$${M_{P^*}^2}-{M_P^2}=-{\textstyle {8\over 3}}
{\tilde G} \EQN sdf1$$
\autoparens
\offparens
$$3{M_{P^*}^2}+{M_P^2}=4({m_Q}+{\bar\Lambda})^2+8{\tilde K}
\EQN sdf2$$
\autoparens
in the heavy quark limit. Heavy quark symmetry constrains the
combination \Eq{sdf2} to be independent of the mass-squared splitting,
\Eq{sdf1}. Of course, it follows from
\Eq{sdf2} that the $D$-${D^*}$ and $B$-${B^*}$ mass-squared splittings
are equal in the heavy quark limit since ${\tilde G}$ is independent
of the heavy quark mass.

If ${P}$ and ${P^*}$ are paired in the sense of \Eq{basicmultst}, the
masses must be related as ${\mu^2}\pm\delta$. But this would violate
the heavy quark symmetry constraints on the mass-squared matrix.
In order to reconcile
the chiral constraint with the heavy quark symmetry constraints of
\Eq{sdf1} and \Eq{sdf2} we 
must introduce additional heavy meson states. The first excited heavy
mesons (not yet observed) have ${s_{\ell}}={\oneh}$ and ${\pi
_{\ell}}=(+)$, and are denoted $P_0^*$ ($0^+$) and $P_1'$ ($1^+$).
The unique solution to the combined chiral and heavy quark constraints
is then\ref{silas}

\offparens
$$\eqalign{
&{M_P^2}={\mu^2}-{\textstyle {3\over 2}}\epsilon\qquad{M_{P^*}^2}
        ={\mu^2}+{\textstyle {1\over 2}}\epsilon\cr
&{M_{P_0^*}^2}={\mu^2}+{\textstyle {3\over 2}}\epsilon\qquad {M_{P_1'}^2}=
{\mu^2}-{\textstyle {1\over 2}}\epsilon\cr}
\EQN phymassesavw1 $$\autoparens
where $\mu^2=({m_Q}+{\bar\Lambda})^2+2{\tilde K}$ $\in\;{{\hat
M}_0^2}$ and $\epsilon=-4{\tilde G}/3$ $\in\;{{\hat M}_{\sss
\vev{{\bar q}q}}^2}$.  Hence $P$ and ${P_0^*}$ are paired and ${P_1'}$
and ${P^*}$ are paired. What representation of $SU(2)\times SU(2)$
does this solution correspond to? We consider two scenarios.
Scenario {\bf\underbar{a}} corresponds to naive pairing of the meson states 
consistent with \Eq{basicmultst}:

\vskip0.1in
\noindent{\bf\underbar{a}}:
\vskip0.1in

\offparens
$$\eqalign{
&\ket{P}=\ket{\psi_-}_{\sss 1}\qquad\;\;
\ket{P^*}^{\sss (0)}
=\ket{\psi_+}_{\sss 2} \cr 
&\ket{P_0^*}=\ket{\psi_+}_{\sss 1}\qquad
\ket{P_1'}^{\sss (0)}
=\ket{\psi_-}_{\sss 2} \cr}
\EQN chstates1pre $$\autoparens
where

\offparens
$$
\ket{\psi_\pm}_i={\onehtsq}\{\ket{0\; {\oneht}\; Q}_{\sss i}
\pm\ket{{\oneht}\; 0\; Q}_{\sss i}\}. 
\EQN ches2 $$\autoparens
The subscripts denote distinct states.  The symbol $Q$ represents
quantum numbers carried by the heavy quark. With this representation
content we have

\offparens
$$\eqalign{
&{M_P^2}={{\bar\mu}^2}-{\bar\epsilon}\qquad
{M_{P^*}^2}={\nu^2}+\kappa \cr
&{M_{P_0^*}^2}={{\bar\mu}^2}+{\bar\epsilon}
\qquad {M_{P_1'}^2}={\nu^2}-\kappa \cr}
\EQN phymassesavw2pre $$\autoparens
where we have defined

$$
{_{\sss 1}}\bra{{\oneht}\; 0\; Q}{{\hat M}_{\sss \vev{{\bar q}q}}^2} 
\ket{0\;{\oneht}\; Q}_{\sss 1}=
{\bar\epsilon}\qquad
{_{\sss 2}}\bra{{\oneht}\; 0\; Q}{{\hat M}_{\sss \vev{{\bar q}q}}^2} 
\ket{0\; {\oneht}\; Q}_{\sss 2}=
{\kappa}
\EQN funds3pre $$ 
$$
{_{\sss 1}}\bra{{\oneht}\; 0\; Q}{{\hat M}_0^2} 
\ket{{\oneht}\; 0\; Q}_{\sss 1}=
{\bar\mu}^2\qquad
{_{\sss 2}}\bra{{\oneht}\; 0\; Q}{{\hat M}_0^2} 
\ket{{\oneht}\; 0\; Q}_{\sss 2}=
{\bar\nu}^2.
\EQN genfunds67pre $$ 
In the heavy quark limit the solution of \Eq{phymassesavw1} is
recovered if ${\bar\mu}={\bar\nu}=\mu$, ${\bar\epsilon}=3\epsilon/2$ and
$\kappa =\epsilon/2$. We then have

$$
{_{\sss 1}}\bra{{\oneht}\; 0}{{\hat M}_{\sss \vev{{\bar q}q}}^2} 
\ket{0\;{\oneht}}_{\sss 1}=3\;
{_{\sss 2}}\bra{{\oneht}\; 0}{{\hat M}_{\sss \vev{{\bar q}q}}^2} 
\ket{0\;{\oneht}}_{\sss 2}
\EQN contra$$
where we have removed the $Q$ label since these matrix elements are
independent of heavy quark properties. This equation is clearly
incompatible with the universality conjecture.  We therefore rule out
scenario {\bf\underbar{a}} as a viable chiral representation of the
heavy mesons.

As pointed out above, we can instead make combinations of
\Eq{basicmultst} which give {\it identical} predictions for the
observable mass-squared splittings and pion transitions and yet which
order matrix elements differently:

\vskip0.1in
\noindent{\bf\underbar{b}}:
\vskip0.1in

\offparens
$$\eqalign{
&\ket{P}={\onehtsq}\{ \ket{\psi_-}_{\sss 1} + \ket{\psi_-}_{\sss 2} \}\qquad
\quad\ket{P^*}^{\sss (0)}
={\onehtsq}\{ \ket{\psi_+}_{\sss 1} - \ket{\psi_+}_{\sss 2} \}\cr 
&\ket{P_0^*}=-{\onehtsq}\{ \ket{\psi_+}_{\sss 1} 
+ \ket{\psi_+}_{\sss 2} \}\qquad
\ket{P_1'}^{\sss (0)}
=-{\onehtsq}\{ \ket{\psi_-}_{\sss 1} - \ket{\psi_-}_{\sss 2} \}.\cr}
\EQN chstates1 $$\autoparens
We then have

\offparens
$$\eqalign{
&{M_P^2}=({{\bar\mu}^2}+{\nu^2})-({\bar\epsilon} +\kappa )\qquad
{M_{P^*}^2}=({{\bar\mu}^2}-{\nu^2})+({\bar\epsilon} -\kappa )\cr
&{M_{P_0^*}^2}=({{\bar\mu}^2}+{\nu^2})+({\bar\epsilon} +\kappa )
\qquad {M_{P_1'}^2}=({{\bar\mu}^2}-{\nu^2})-({\bar\epsilon} -\kappa )\cr}
\EQN phymassesavw2 $$\autoparens
where we have defined

$${_i}\bra{{\oneht}\; 0\; Q}{{\hat M}_{\sss \vev{{\bar q}q}}^2} 
\ket{0\;{\oneht}\; Q}_j=
{\delta_{ij}}{\bar\epsilon}+(1-{\delta_{ij}}){\kappa}
\EQN funds3 $$ 
$${_i}\bra{{\oneht}\; 0\; Q}{{\hat M}_0^2} \ket{{\oneht}\; 0\; Q}_j
={\delta_{ij}}{{\bar\mu}^2}+(1-{\delta_{ij}}){\nu^2}.
\EQN genfunds67 $$ 
In the heavy quark limit the solution of \Eq{phymassesavw1} is
recovered if ${\bar\mu}=\mu$, ${\nu}=0$, ${\bar\epsilon}=\epsilon$ and
$\kappa =\epsilon/2$.  We then have

$${_i}\bra{{\oneht}\; 0\; Q}{{\hat M}_{\sss \vev{{\bar q}q}}^2} 
\ket{0\;{\oneht}\; Q}_j=
{_i}\bra{{\oneht}\; 0}{{\hat M}_{\sss \vev{{\bar q}q}}^2} 
\ket{0\;{\oneht}}_j=
\oneht({\delta_{ij}}+1)\epsilon
\EQN funds3mod $$ 
$${_i}\bra{{\oneht}\; 0\; Q}{{\hat M}_0^2} \ket{{\oneht}\; 0\; Q}_j
={\delta_{ij}}{{\mu}^2},
\EQN genfunds67mod $$ 
where in \Eq{funds3mod} we have made use of the fact that $\epsilon$
is independent of heavy quark properties. This assignment of states is
consistent with universality. Moreover, one can easily check that
universality is not inconsistent with $1/m$ corrections.  Therefore
scenario {\bf\underbar{b}} is a viable chiral representation.  We have

$$
{M_{P_0^*}^2}- {M_{P_1'}^2}={M_{P^*}^2}-{M_P^2}=
2\,{_{\sss 1}}\bra{{\oneht}\; 0}{{\hat M}_{\sss \vev{{\bar q}q}}^2}
\ket{0\;{\oneht}}{_{\sss 1}} =
2\,{_{\sss 2}}\bra{{\oneht}\; 0}{{\hat M}_{\sss \vev{{\bar q}q}}^2}
\ket{0\;{\oneht}}{_{\sss 2}}.
\EQN barco$$
On the basis of the universality assumption of \Eq{kmr234} we then
find

$$
{M_{P^*}^2}-{M_P^2}=\Delta ={M_{K^*}^2}-{M_K^2}
\EQN swindle$$
where the last equality follows from \Eq{kmr2}.
\Eq{swindle} is compared with experiment in Table 1.
This derivation is rendered less persuasive by the universality
conjecture. Nevertheless, it is the best that we can do. We will give
an independent test of universality in the next section.

\vskip0.1in
\noindent {\twelvepoint{\bf 5.\quad A Note on the Baryons}}
\vskip0.1in

The excited $I=1/2$ cascade $\Xi(1530)$ decays to $\Xi$ and a pion
with a branching ratio of one\ref{pdg}. We therefore expect these
states to be paired in the sense of \Eq{basicmultst}.  Hence we have
another test of universality. We have

$$\eqalign{
&\ket{\Xi (1530)}={1\over\sqrt{2}}\{\ket{0\; {\oneht}\; 2s}
           +\ket{{\oneht}\; 0\; 2s}\}\cr 
&\ket{\Xi}={1\over\sqrt{2}}\{\ket{0\; {\oneht}\; 2s}
           -\ket{{\oneht}\; 0\; 2s}\}\cr}
\EQN chstates4 $$\autoparens
where $2s$ denotes the strange quark content.
It follows that

$${M_{\Xi (1530)}^2}-{M_{\Xi}^2}= 
2\bra{{\oneht}\; 0\; 2s}{{\hat M}_{\sss \vev{{\bar q}q}}^2}
\ket{0\;{\oneht}\; 2s} =\Delta \EQN kmr2$$ 
where the last equality follows from the universality assumption of
\Eq{kmr234}. This predicts a ${\Xi (1530) -\Xi }$ mass-squared splitting 
of $0.50$ in units of $1/{\alpha '}$, as compared to the observed
splitting of $0.54$.  Of the baryon pairs listed in Table 5, this
splitting is in fact closest to the $\rho$-$\pi$ splitting, thus
providing a gratifying test of the universality conjecture.

Other baryons require separate discussion. For instance, the $I=1/2$
baryons made out of three light quarks clearly have single-pion
transitions to states with $I=3/2$. So states like $N$ and $\Delta$
will in general fill our reducible combinations of any number of
$(0,\oneh )$, $(\oneh ,0 )$, $(0,\threeh )$, $(\threeh ,0)$, $(1,\oneh
)$, and $(\oneh , 1)$ irreducible representations of ${SU(2)}\times
{SU(2)}$\ref{mended}. Many possible $\discrete$ multiplets can be
constructed consistent with the allowed mass-squared splittings. One
might further expect that the $I=0$ and $I=1$ baryons have a quartet
structure similar to that of the light mesons.

\table{avw2}
\caption{Lowest lying baryons of a given character. Masses are
central values from the particle data group, and 
$\alpha ' ={0.88}\;{{GeV}^{-2}}$.}
\ruledtable
$A^*$-$A$ | ${\alpha '}({M_{A^*}^2}-{M_{A}^2})$ \cr $\Delta -N$ |
$0.56$\crnorule $\Sigma (1385) -\Lambda$ | $0.59$\crnorule $\Sigma
(1385) -\Sigma $ | $0.44$\crnorule $\Xi (1530) -\Xi $ |
$0.54$\crnorule $\Sigma_c -\Lambda_c $ | $\;0.70$
\endruledtable
\endtable

\vskip0.1in
\noindent {\twelvepoint{\bf 6.\quad Regge Behavior and QCD}}
\vskip0.1in

It might seem mysterious that using the full chiral algebra and
$\discrete$ we recovered results that are predicted by hadronic string
models. Consider, however, that {\it the basic assumptions that have
been made in this paper are in one-to-one correspondence with
statements of Regge asymptotic behavior}. This correspondence is
exhibited in Table 6 and was demonstrated long ago by Weinberg in
\Ref{alg} and further developed in \Ref{mended}. The constraints
on the mass-squared matrix are known as superconvergent sum
rules. Hadronic string models exhibit very soft asymptotic behavior;
in fact, they satisfy an infinite number of superconvergence
relations\ref{cley}.  We required only two superconvergence relations
to derive equal spacing relations for the low-lying mesons.

In \Ref{mended} Weinberg showed that the algebraic relation $[{{\hat
M}_0^2},{{\hat M}_{\sss\vev{{\bar q}q}}^2}]=0$, which is a statement
about diffraction in pion-hadron scattering (see Table 6), fixes the
reducible chiral representations filled out by light mesons. Following
\Ref{beane}, in this paper we instead used a $\discrete$ symmetry to fix 
the chiral representations. That these statements are equivalent is
easy to see. We have seen that the $\discrete$ transformation acts as
${{\hat M}_0^2}\leftrightarrow{{\hat M}_{\sss \vev{{\bar q}q}}^2}$ on
the mass-squared matrix (see Section 2.2). This transformation is a
symmetry if $[{{\hat M}_0^2},{{\hat M}_{\sss\vev{{\bar q}q}}^2}]=
-[{{\hat M}_0^2},{{\hat M}_{\sss\vev{{\bar q}q}}^2}]$ which holds if
and only if $[{{\hat M}_0^2},{{\hat M}_{\sss\vev{{\bar q}q}}^2}]=0$.

\table{analog}
\caption{ 
The equivalence of the first and second columns is exact in the
tree-graph approximation (large-$\nc$ for pion-meson scattering). The
first row implies that, for each helicity, mass eigenstates fill out
generally reducible representations of $SU(2)_L\times SU(2)_R$.
That the two parts of ${{\hat M}^2}$ commute is a statement of
the $\discrete$ symmetry.}
\doublespaced
\ruledtable
Hadrons   | Regge in $\pi\alpha\rightarrow\pi\beta$   \cr
$SU({2})_L\times SU({2})_R$|${\alpha_1} (0) < 1$ \cr
${{\hat M}^2}=
{{\hat M}_0^2}+{{\hat M}_{\sss \vev{{\bar q}q}}^2}$ |${\alpha_2} (0)<0$, \quad ${\alpha_0} (0)=1$\cr
${\bf Z_2}\leftrightarrow$
$[{{\hat M}_0^2},{{\hat M}_{\sss \vev{{\bar q}q}}^2}]=0$|${\alpha_0} (0)<0\quad \alpha\neq\beta$
\endruledtable
\endtable

Here we suggest a QCD-based interpretation of the $\discrete$
symmetry. Consider an underlying quark description with an
$SU(2)\times SU(2)$ flavor symmetry and the desired $\discrete$
symmetry.  Assume that the quarks transform as $(0,{{\oneh}})$ and
$({{\oneh}},0)$ with respect to $SU(2)\times SU(2)$. The $\discrete$
transformation permutes quarks with opposite chiral charges: i.e.,
$(0,{{\oneh}})\leftrightarrow ({{\oneh}},0)$. Hence the permutation
properties of meson states assumed in this paper. It is important to
realize that this symmetry is {\it not} ordinary parity; the
$\discrete$ symmetry is an internal symmetry, not a spacetime
symmetry. Therefore, the $\discrete$ symmetry must act independently
on left- and right-handed quarks. For instance, a left-handed Weyl
fermion of charge $(0,{{\oneh}})$ must have a left-handed partner of
charge $({{\oneh}},0)$.  If parity is conserved, then the quark
description must have equal numbers of left- and right-handed quarks
assigned to each chiral charge, $(0,{{\oneh}})$ and
$({{\oneh}},0)$. This implies that the $SU(2)\times SU(2)$ flavor
symmetry in the underlying quark description with the $\discrete$
symmetry must be {\it vectorlike}\ref{nn}. This is easy to see by
noticing that $SU(2)\times SU(2)$ invariant mass operators can always
be formed when $\discrete$ is a symmetry\ref{silas}.

Clearly the $\discrete$ symmetry is not a symmetry of the QCD
lagrangian in which the $SU(2)\times SU(2)$ flavor symmetry is {\it
chiral}. But what then do we learn from the phenomenological successes
of the $\discrete$ symmetry? Consider the following argument. An
underlying quark description with the $\discrete$ symmetry is
vectorlike and is therefore automatically consistent with the
Nielsen-Ninomiya theorem\ref{nn}. This implies that the quark
description can be defined on the lattice with an {\it exact}
$SU(2)\times SU(2)$ symmetry at non-zero lattice spacing. In practice
such a description will have four flavors of quarks and so will look
like QCD with doublers\ref{silas}. The origin of the doublers has a
physical interpretation in terms of chiral anomalies\ref{ks}. Since a
gauge theory with an intrinsic cutoff cannot feel the effects of
anomalies, each left- and right-handed Weyl fermion in the theory will
have a doubler of opposite chiral charge. However, only for special
values of the vacuum parameters of the theory will there be a
$\discrete$ symmetry corresponding to permutations of these charges.
For instance, the $\discrete$ symmetry can be spontaneously broken by
quark condensates. This would correspond to a low-energy theory with
$[{{\hat M}_0^2},{{\hat M}_{\sss\vev{{\bar q}q}}^2}]\neq 0$.  It is
intriguing that the $\discrete$ symmetry is also broken {\it unless}
$\bar\theta$ takes $CP$ conserving values
\ref{silas}\ref{sb2}. 

It is surprising that this $\discrete$ symmetry has physical
consequences relevant to low-energy QCD.  A possible explanation is
that the $\discrete$ symmetry will play a role in any nonperturbative
definition of QCD where the flavor symmetries are unbroken by the
regulator$^4$\vfootnote4{This viewpoint is elaborated in
\Ref{sb3}.}. Since the lattice is the only known means of defining QCD
in the nonperturbative region, and no lattice definition exists in
which the flavor symmetries are chiral and unbroken, this hypothesis
is safe. In essence, this hypothesis raises the Nielsen-Ninomiya
theorem from a statement specific to the lattice to the level of a
general physical principle.

\figure{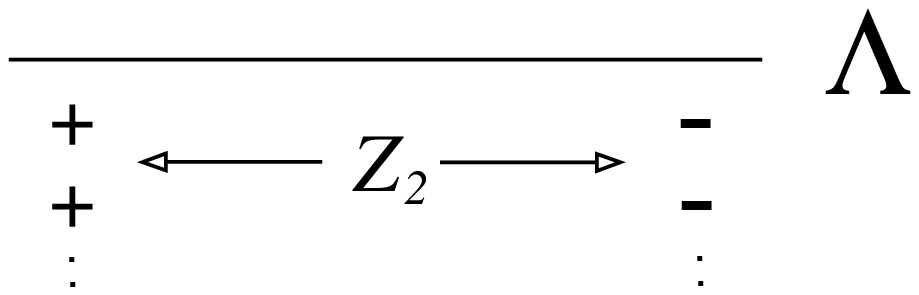}
\epsfxsize 3in
\centerline{\epsfbox{mirror2.eps}}
\caption{The anomaly as a quantum mirror. A theory with an
intrinsic cutoff does not feel the effects of anomalies.  Each Weyl
fermion has a partner of opposite chiral charge. For instance, each
quark transforming as $(0,{{\oneh}})$ (+) has a partner with charge
$({{\oneh}},0)$ (-). For special values of vacuum parameters there is
an invariance with respect to a $\discrete$ permutation of these
charges.}
\endfigure

There is some evidence backing our hypothesis.  The equal-spacing
relations have been tested in lattice QCD using improved lattice
actions\ref{wolo}. Improved actions do substantially better at
generating universal (constant) mass-squared splittings than do Wilson
or staggered fermion actions. In the Wilson action the chiral symmetry
breaking Wilson term is dimension five whereas in the improved action
it is dimension seven\ref{wolo}. It is conceivable that only an action
improved to all orders will explicitly realize the $\discrete$
symmetry and give precisely constant mass-squared splittings.

\vskip0.1in
\noindent {\twelvepoint{\bf 7.\quad Summary and Conclusion}}
\vskip0.1in

There has been little progress in understanding empirically successful
predictions of Regge theory and string models of hadrons from QCD. One
piece of phenomenology that has received little attention is the
remarkable equality of various hadron mass-squared splittings 
predicted long ago by hadronic string models.

In this paper we have used symmetry arguments to derive a cornucopia
of equal spacing relations. Our arguments rest on consequences of the
full chiral algebra with two and three flavors of quarks, together
with a $\discrete$ symmetry which permutes representations of the
chiral algebra. We showed that ${M_{\rho}^{2}}-{M_{\pi}^{2}} =
{M_{\sss {K*}}^{2}}-{M_{\sss K}^{2}}$ to leading order in $m_q$. This
and other equal spacing relations for the masses of the low-lying
pseudoscalar, vector, scalar and axialvector octets are our most
robust results. We also considered isospin violation and gave new
determinations of ratios of current quark masses.  The requirement
that all states fill out representations of $SU(2)\times SU(2)$ {\it
and} $SU(3)\times SU(3)$ gave several interesting results.  It enabled
us to determine members of the scalar octet that are ambiguous from
the point of view of $SU(3)$ alone and it required that there be
nontrivial octet-singlet mixings. It further allowed us to express the
kaon mass-squared splittings in terms of an $I=1/2$ matrix element
which is independent of strange quark properties.  We conjectured that
this matrix element is universal. We then found the chiral
representations of the heavy mesons that are consistent with
$\discrete$ and heavy quark symmetry and showed that the relevant
mass-squared splittings are determined by the universal matrix
element.  This result gave ${M_{\sss {K*}}^{2}}-{M_{\sss K}^{2}} =
{M_{\sss {P*}}^{2}}-{M_{\sss P}^{2}}$ where $P$ represents $D$ or
$B$. Unfortunately, the necessity of conjecturing the existence of a
universal matrix element renders this derivation less persuasive than
the derivation of the light-meson mass-squared splittings. We then
discussed the baryons and gave an independent successful test of the
universality conjecture using the cascades. Finally, we compared our
derivation to that of hadronic string models and gave an
interpretation of the $\discrete$ symmetry in the context of lattice
QCD. We suggested that a $\discrete$ permutation of chiral charges is
a fundamental property of a gauge theory with an intrinsic cutoff.

\vskip0.15in


This work was supported by the U.S. Department of Energy (Grant
DE-FG05-90ER40592 at Duke and grant DE-FG02-93ER-40762 at Maryland). I
thank B.~M\" uller and R.M.~Woloshyn for useful conversations.

\vfill\eject                                     
\nosechead{References}
\ListReferences
\vfill\supereject

\end